\documentclass{article}
\input{epsf}
\usepackage{amssymb}
\input epsf
\title{\bf Can the Chaplygin gas be a plausible model for dark energy?}
\author{Vittorio Gorini,$^{1,2}$ Alexander Kamenshchik$^{1,3}$ and Ugo
Moschella$^{1,2}$\\[20pt]
$^1$Dipartimento di Scienze Matematiche, Fisiche e
Chimiche, Universit\`a dell'Insubria, \\
Via Valleggio 11, 22100 Como\\
$^2$INFN sez. di Milano, Italy \\
$^3$L.D. Landau Institute for Theoretical Physics of Russian
Academy of Sciences, \\ Kosygin str. 2, 117334, Moscow, Russia}
\date{}
\begin{document}
\maketitle

\begin{abstract}
In this note two cosmological models representing the flat
Friedmann Universe filled with a Chaplygin fluid, with or without
dust, are analyzed in terms of the recently proposed
"statefinder" parameters \cite{statefinder}.  Trajectories of both
models in the parameter plane are shown to be significantly
different w.r.t. "quiessence" and "tracker" models.  The
generalized Chaplygin gas model with an equation of state of the
form $p = -A/\rho^{\alpha}$
is also analyzed in terms of the statefinder parameters.
\end{abstract}

In the search for cosmological models describing the observed
cosmic acceleration \cite{accel1,accel2,accel3}, the inspiration coming from inflation has
suggested mainly models making use of scalar fields
\cite{Star-Varun,Varun,Star,Star1,Sazhin}. There are of course alternatives; in
particular, in \cite{we,Bilic,Fabris} an elementary model has
been presented describing a Friedmann universe filled with a
perfect fluid obeying the Chaplygin equation of state
\begin{equation}
p = -\frac{A}{\rho},
\label{Chapl}
\end{equation}
where $A$ is a positive constant (for a thorough review see Ref.
\cite{Jackiw}). The interesting feature of this model is that it
naturally provides a universe that undergoes a transition from a
decelerating phase, driven by dust-like matter, to a cosmic
acceleration at later stages of its evolution (see \cite{we} for
details). An interesting attempt to justify this model
\cite{Bilic1} makes use of an effective field theory for a
three-brane universe \cite{Sundrum}.

In the flat case, the model can be equivalently described in
terms of a homogeneous minimally coupled scalar field $\phi$, with
potential \cite{we}
\begin{equation}
V(\phi) = \frac{1}{2}\sqrt{A}\left(\cosh 3\phi + \frac{1}{\cosh
3\phi}\right). \label{potential}
\end{equation}
However, since  models trying to provide a description (if not an
explanation) of the cosmic acceleration are proliferating, there
exists the problem of discriminating between the various
contenders. To this aim a new proposal introduced in
\cite{statefinder} makes use of a pair of parameters $\{r,s\}$,
called ``statefinder''. The relevant definition is as follows:
\begin{equation}
r \equiv \frac{\stackrel{\cdots}{a}}{a H^3},\ \ s \equiv \frac{r-1}{3(q-1/2)},
\label{statefinder}
\end{equation}
where $ H \equiv \frac{\dot{a}}{a}$ is the Hubble constant  and $
q \equiv -\frac{\ddot{a}}{a H^2} $ is the deceleration parameter.
The new feature of the statefinder is that it involves the third
derivative of the cosmological radius.

Trajectories in the $\{s,r\}$-plane corresponding to different
cosmological models exhibit qualitatively different behaviors.
$\Lambda$CDM model diagrams correspond to the fixed point $s=0$,
$r=1$. The so-called "quiessence" models \cite{statefinder} are
described by vertical segments with $r$ decreasing from $r=1$
down to some definite value. Tracker models \cite{tracker} have
typical trajectories similar to arcs of parabola lying in the
positive quadrant with positive second derivative.

The current location of the parameters $s$ and $r$ in these
diagrams can be calculated in models (given the deceleration
parameter); it may also be extracted from data coming from SNAP
(SuperNovae Acceleration Probe)-type experiments
\cite{statefinder}. Therefore, the statefinder diagnostic combined
with future SNAP observations may possibly be used to discriminate
between different dark energy models.

In this short paper we apply that diagnostic to Chaplygin
cosmological models (the direct comparison of the already
available supernova data with the Chaplygin gas model was also
undertaken recently in \cite{Fabris1,Avelino}). We consider both
the one-fluid pure Chaplygin gas model and a two-fluid model
where dust is present as well. We show that these models are
different from those considered in \cite{statefinder} and they are
worth of further study.

To begin with, let us rewrite the formulae for the statefinder
parameters \cite{statefinder} in a form convenient for our
purposes. We shall need the Friedmann equation for the flat
universe
\begin{equation}
H^2 = \frac{\dot{a}^2}{a^2} = \rho
\label{Friedmann}
\end{equation}
and the energy conservation equation
\begin{equation}
\dot{\rho} = - 3 H (\rho + p).
\label{Friedmann1}
\end{equation}
Using these two equations it is easy to find that
\begin{equation}
q =  \frac 12+ \frac{3}{2}\frac p\rho, \label{q-find}
\end{equation}
and then
\begin{equation}
r = 1 -  \frac{ 3\dot{p}}{2 \, \rho \sqrt{ \rho}}\, ,
\label{s-find} \;\;\;\; s = -\frac{\dot{p}}{3 p \, \sqrt{\rho}}.
\end{equation}
For a one-component fluid\footnote{We confine ourselves to the
case, which is fulfilled in our models, of a fluid for which the
equation of state has the form $p = p(\rho)$.} these formulae
become especially simple. Since
\begin{equation}
\dot{p} = \frac{\partial p}{\partial \rho}\, \dot{\rho} =
-3\sqrt{\rho}\, (\rho + p) \frac{\partial p}{\partial \rho}\, ,
\label{pressure1}
\end{equation}
we easily get:
\begin{equation}
r = 1 + \frac{9}{2}\left(1 + \frac{p}{\rho}\right) \frac{\partial
p}{\partial \rho} \label{s-find1}\, , \;\;\;\; s = \left(1 +
\frac{\rho}{p}\right) \frac{\partial p}{\partial \rho}\, .
\end{equation}
For the Chaplygin gas one has simply that
\begin{equation}
v^2_s = \frac{\partial p}{\partial \rho}= \frac{A}{\rho^2} =
-\frac{p}{\rho} = 1+s \label{x-define}
\end{equation}
($v^2_s$ is the square of the velocity of sound) and therefore
\begin{equation}
r = 1 - \frac{9}{2}\, s\, (1+s). \label{r-find2}
\end{equation}
Thus, the curve of $r(s)$ is an arc of parabola. To find the
admissible values of $s$, we note that Eqs. (\ref{Chapl}),
(\ref{Friedmann}) and (\ref{Friedmann1}) easily give the
following dependence of the energy density on the cosmological
scale factor \cite{we}:
\begin{equation}
\rho = \sqrt{A + \frac{B}{a^6}}, \label{Chapl1}
\end{equation}
where $B$ is an integration constant; therefore
\begin{equation}
v^2_s = 1+s=\frac{A}{A + \frac{B}{a^6}}. \label{x-define0}
\end{equation}
When the cosmological scale factor $a$ changes from $0$ to
$\infty$ the velocity of sound varies from $0$ to $1$ and $s$
varies from $-1$ to $0$. Thus in our model the statefinder $s$
takes negative values; this feature is not shared by quiessence
and tracker models considered in \cite{statefinder}.

\begin{figure}[h]
  \centering
  \epsfxsize 12cm \epsfbox{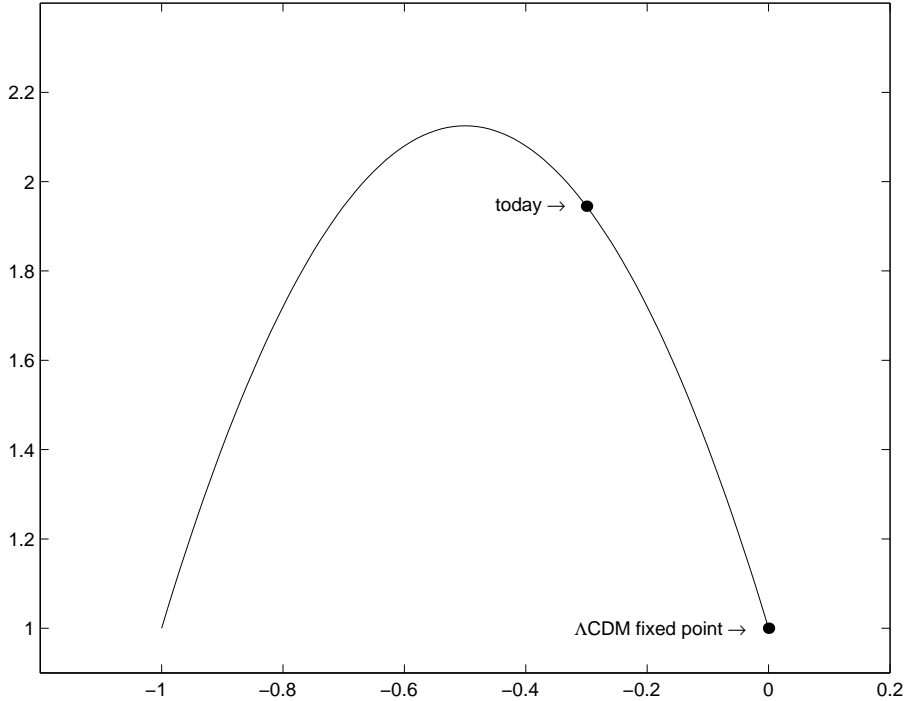}
  \caption{s-r evolution diagram for the pure Chaplygin gas}\label{Fig.1}
\end{figure}

As $s$ varies in the interval $[-1,0]$, $r$ first increases from
$r=1$ to its maximum value and then decreases  to the
$\Lambda$CDM fixed point $s=0$, $r=1$ (see Fig. \ref{Fig.1}).

\noindent If $q \approx - 0.5$ the current values of the
statefinder (within our model) are $ s \approx -0.3,\ \ r \approx
1.9 $. In \cite{statefinder} it is reported an interesting
numerical experiment based on 1000 realizations of a SNAP-type
experiment, probing a fiducial $\Lambda$CDM model. Our values of
the statefinder lie outside the three-sigma confidence region
displayed in \cite{statefinder}. Based on this fact it can be
expected that future SNAP experiments should be able to
discriminate between the pure Chaplygin gas model and the standard
$\Lambda$CDM model.

\begin{figure}[h]
  \centering
  \epsfbox{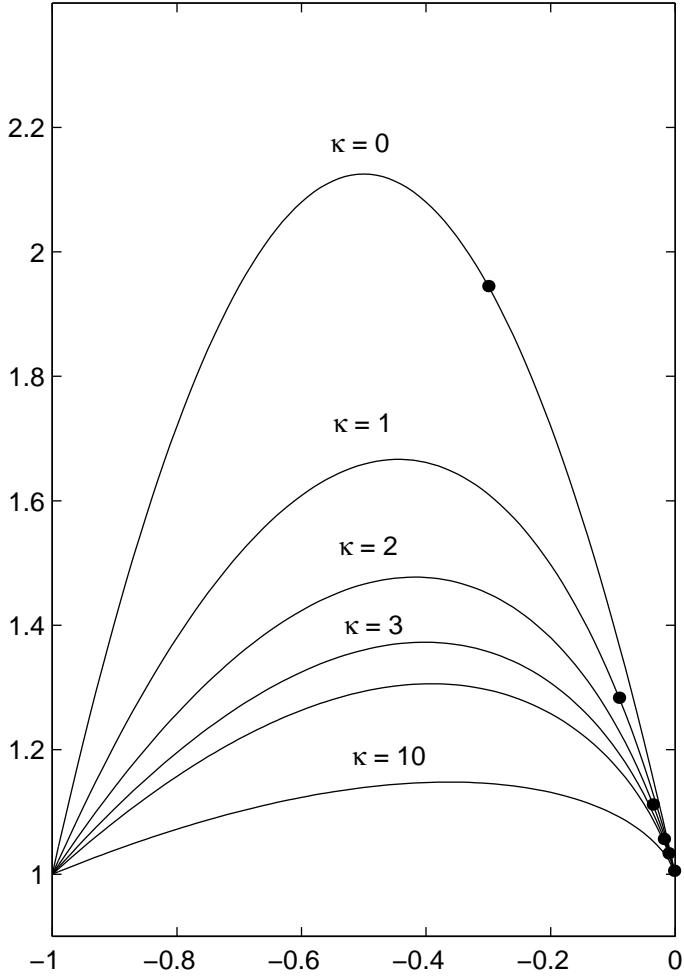}
  \caption{s-r evolution diagram for the Chaplygin gas mixed with dust.
  Dots locate the current value of  the statefinder}\label{Fig.2}
\end{figure}

Let us consider now a more "realistic" cosmological model which,
besides a Chaplygin's component, contains also a dust component.
For a two-component fluid Eqs. (\ref{s-find}) take the following
form:
\begin{equation}
r = 1 + \frac{9}{2({\rho+ \rho_1})}\left[{ \frac{\partial
p}{\partial \rho}(\rho + p)+\frac{\partial p_1}{\partial
\rho_1}(\rho_1 + p_1) }\right], \label{r-find3}
\end{equation}
\begin{equation}
s = \frac{1}{p + p_1} \left[\frac{\partial p}{\partial \rho}(\rho
+ p) + \frac{\partial p_1}{\partial \rho_1}(\rho_1 + p_1)\right].
\label{s-find3}
\end{equation}
If one of the fluids is dust, i.e. $p_1 = p_d = 0$,the above formulae become
\begin{equation}
r = 1 + \frac{9(\rho + p)}{2(\rho + \rho_d)} \frac{\partial
p}{\partial \rho}\, ,\;\;\;\; s = \frac{\rho + p}{p}
\,\frac{\partial p}{\partial \rho}. \label{s-find4}
\end{equation}
If the second fluid is the Chaplygin gas, proceeding exactly as
before we obtain the following relation:
\begin{equation} r = 1 -
\frac{9}{2}\frac{s(s+1)}{1 + \frac{\rho_d}{\rho}}. \label{r-find5}
\end{equation}
To find the term $\rho_d/\rho$ we write down the dependence of the
dust density on the cosmological scale factor:
\begin{equation}
\rho_d = \frac{C}{a^3}, \label{dust1}
\end{equation}
where $C$ is a positive constant. Eq. (\ref{x-define0}) gives that
$A a^6 + B = -\frac{B}{s}$ and therefore
\begin{equation}
\frac{\rho_d}{\rho} = \frac{C}{\sqrt{A a^6 + B}} =  \kappa
\sqrt{-s} , \label{dens-rel2}
\end{equation}
where the constant $ \kappa = {C}/{\sqrt{B}} $ is the ratio
between the energy densities of dust and of the Chaplygin gas at
the beginning of the cosmological evolution (cf. Eq.
(\ref{Chapl1})). Thus
\begin{equation}
r = 1 - \frac{9}{2}\frac{s(s+1)}{1 + \kappa\sqrt{-s}}.
\label{r-find6}
\end{equation}
Graphs of the function (\ref{r-find6}) for different choices of
$\kappa$ are plotted in Fig. \ref{Fig.2}.

In this case there are choices of the parameters so that the
current values of the statefinder are close to the $\Lambda$CDM
fixed point; indeed for $\kappa = 1$ we have $s=-0.09$ and
$r=1.2835$. By increasing $\kappa$ we get closer and closer.
Already for $\kappa = 2$ we get $s=0.035$, $r=1.11$ while for
$\kappa > 10$ the statefinder essentially coincides with the
$\Lambda$CDM fixed point (see Fig. \ref{Fig.2}).

Thus our  two-fluid cosmological models (with  $\kappa$ say
bigger than 2) cannot be discriminated from the $\Lambda$CDM model
on the basis of the statefinder analysis. On the other hand this
fact could also be interpreted as an argument in favor of the
model because the $\Lambda$CDM model does not contradict
observations up to now.

The attractive point of our two-fluid cosmological model with
a parameter $\kappa$ of order one is that it may suggest a solution to
the cosmic coincidence
conundrum: indeed here the initial values of the energies of dust
and of the Chaplygin gas are of the same order of magnitude.

The comparison of the two-fluid (i.e the Chaplygin gas plus dust)
cosmological model with observational data has also been studied in
\cite{Fabris1,Avelino}. An analysis of the data obtained from 26
Supernovae leads to the conclusion \cite{Fabris1} that the best
fitting model seems to be the pure Chaplygin gas without dust. On
the other hand \cite{Avelino} by combining the results of the
observations of 92 Supernovae with the matter power spectrum a
wide range of  values of $\kappa$ seems to be admissible.

Thus, for getting more precise constraints on the parameters of
the Chaplygin models one needs both new observations and additional
diagnostic techniques. From this point of view an application of the
statefinder analysis could be useful.

Similar conclusions also hold for a one-fluid model of a
generalized Chaplygin gas with a modified equation of state, as
introduced in \cite{we}:
\begin{equation}
p = -\frac{A}{\rho^{\alpha}},
\label{general}
\end{equation}
with $0 \leq \alpha \leq 1$. This gives a cosmological evolution
from an initial dust-like behavior to  an asymptotic cosmological
constant, with an intermediate epoch that can be seen as a
mixture of a cosmological constant with a fluid obeying the state
equation $p = \alpha \rho$ ($\alpha = 0$ corresponds to the $\Lambda$CDM
model). This generalized model was studied in
some detail in  \cite{Bertolami}. Equation (\ref{general}) gives
the following dependence of the energy density $\rho$ on the
scale factor $a$:
\begin{equation}
\rho = \left(A + \frac{B}{a^{3(\alpha+1)}}\right)^{\frac{1}{\alpha+1}}.
\label{general1}
\end{equation}
In this case the squared velocity of sound
\begin{equation}
v^2_s = \frac{\partial p}{\partial \rho} = - \frac{\alpha p}{\rho}
= \frac{A\alpha}{A + \frac{B}{a^{3(\alpha+1)}}} \label{general2}
\end{equation}
varies from $0$ to $\alpha$. From Eq. (\ref{s-find1}) it follows
that $s = v^2_s -\alpha$ and  therefore the admissible values of
$s$ are now in the interval between $-\alpha$ and $0$. From Eq.
(\ref{s-find1}) we get
\begin{equation}
r = 1 - \frac{9}{2} \frac{s(s + \alpha)}{\alpha} \label{present}
\end{equation}
For small values of $\alpha$ the generalized Chaplygin gas model
becomes
indistinguishable from the standard $\Lambda$CDM cosmological
model.

In conclusion. From the statefinder viewpoint both the pure
Chaplygin model and the two-fluid mixture have different behaviors
w.r.t. other commonly studied models. In particular, a future
larger amount of data on high $z$ type Ia supernovae may allow to
distinguish between the pure Chaplygin gas and the $\Lambda$CDM
models. Instead, the mixed two-fluid model becomes practically
indistinguishable from $\Lambda$CDM for sufficiently large values
of the parameter $\kappa$ ($\kappa > \kappa_0 \approx 5$). On the
other hand a mixed two-fluid model with $\kappa$ of order one is
attractive from the point of view of a possible solution of the
cosmic coincidence problem.

A.K. is grateful to CARIPLO Science Foundation and to University
of Insubria for financial support. His work was also partially
supported by the Russian Foundation for Basic Research under the
grants No 02-02-16817 and 00-15-96699. A.K. is grateful to A.A.
Starobinsky for useful discussions.

\end{document}